\documentclass[12pt,letterpaper]{article}
\usepackage{amsbsy,amsmath,amssymb,graphicx}
\usepackage{geometry}
\usepackage{natbib}

\def\eqd{{\buildrel d \over =}}
\def\bSig\mathbf{\Sigma}



\begin{document}
  \newcommand{\VS}{V\&S}
  \newcommand{\tr}{\mbox{tr}}
  \newcommand{\bx}{\boldsymbol{x}}
  \newcommand{\by}{\boldsymbol{y}}
  \newcommand{\bz}{\boldsymbol{z}}
  \newcommand{\bU}{\boldsymbol{U}}
  \newcommand{\bX}{\boldsymbol{X}}
  \newcommand{\bY}{\boldsymbol{Y}}
  \newcommand{\bZ}{\boldsymbol{Z}}
  \newcommand{\bu}{\boldsymbol{u}}
  \newcommand{\bmu}{\boldsymbol{\mu}}
  \newcommand{\bdelta}{\boldsymbol{\delta}}
  \newcommand{\bsigma}{\boldsymbol{\sigma}}
  \newcommand{\bpi}{\boldsymbol{\pi}}
  \newcommand{\btheta}{\boldsymbol{\theta}}
  \newcommand{\bTheta}{\boldsymbol{\Theta}}

\title{PICS: Probabilistic Inference for ChIP-seq}
\author{Xuekui Zhang$^{1}$,
Gordon Robertson$^{2}$,
Martin Krzywinski$^{2}$,
Kaida Ning$^{3}$,\\
Arnaud Droit$^{4}$,
Steven Jones$^{2}$,
and
Raphael Gottardo$^{4,5}$\footnote{To whom correspondance should be addressed \tt{raphael.gottardo@ircm.qc.ca}}\\
$^{1}$Department of Statistics and $^{3}$Bioinformatics Training Program\\
University of British Columbia, Vancouver, BC, Canada\\
$^{2}$BC Cancer Agency Genome Sciences Centre, Vancouver, BC,
Canada\\
$^{4}$Clinical Research Institute of Montreal (IRCM)\\
$^{5}$Universit\'e de Montreal, Montreal, QC, Canada
}

\maketitle

\label{firstpage}

\baselineskip=18pt

\begin{abstract}
ChIP-seq, which combines chromatin immunoprecipitation with massively parallel short-read
sequencing, can profile in vivo genome-wide transcription factor-DNA association with higher
sensitivity, specificity and spatial resolution than ChIP-chip. While it presents new opportunities
for research, ChIP-seq poses new challenges for statistical analysis that derive from the
complexity of the biological systems characterized and the variability and biases in its digital
sequence data. We propose a method called PICS (Probabilistic Inference for ChIP-seq) for
extracting information from ChIP-seq aligned-read data in order to identify regions bound by
transcription factors.
PICS identifies enriched regions by modeling local concentrations of directional reads, and uses
DNA fragment length prior information to discriminate closely adjacent binding events via a
Bayesian hierarchical $t$-mixture model. Its per-event fragment length estimates also allow it
to remove from analysis regions that have atypical lengths. PICS uses
pre-calculated, whole-genome read mappability profiles and a truncated $t$-distribution to adjust
binding event models for reads that are missing due to local genome
repetitiveness. It estimates uncertainties in model parameters that can be used
to define confidence regions on binding event locations and to filter estimates. Finally, PICS calculates
a per-event enrichment score relative to a control sample, and can use a control sample to
estimate a false discovery rate.
We compared PICS to the alternative methods  MACS, QuEST, and CisGenome, using
published GABP and FOXA1 data sets from human cell lines, and found that PICS'
predicted binding sites were more consistent with computationally predicted binding motifs. 

\end{abstract}

\section{Introduction}
\label{s:intro}
ChIP-seq combines chromatin immumoprecipitation with massively parallel short-read sequencing \citep{b515,b516,b503,b517,b1}. 
It offers high specificity, sensitivity and spatial resolution in profiling diverse aspects of
cellular biology: protein-DNA association \citep{b509,b508,b511,b512,b9,b8}
; histones, histone variants and modified histones \citep{b514,b510,b10,b513,b601}; 
DNA methylation
\citep{b602}; polymerases and transcriptional machinery complexes \citep{b506,b508};
and nucleosome positioning
\citep{b14}.
While sequencing overcomes certain limitations of profiling with microarrays (ChIP-chip),
it raises statistical and computational challenges, some of which are related to those
for ChIP-chip, and others that are novel. A typical ChIP-seq data set consist of
millions or tens of millions of sequence reads that are generated from ends of DNA fragments.
Read lengths are currently typically in the range of 36-50 bp, and the quality of called bases
varies along and between reads; as the sequencing technology evolves, read lengths and
quality, and the number of sequence reads generated in a machine run are increasing. While
pairs of end reads can be generated from each DNA fragment, current ChIP-seq data
typically consist of single-end reads, in which each immunoprecipitated DNA fragment
contributes a directional read from only one randomly selected fragment end.

After read sequences have been aligned to a reference genome \citep{b516},
the goal of subsequent analysis is to transform the aligned read data into a form
that reflects the local density of immunoprecipitated DNA fragments, and, in the
work described here, to estimate locations where transcription factors were associated with
DNA in the experimental cellular system. Analysis is complicated by biases in local read densities
that can be introduced by sequencing and aligning, and by chromatin structure
and genome copy number variations \citep{b506,b516,b503,b505,b530}. As well, 
%when reads from immunoprecipitated DNA fragments overlap repetitive genomic regions, the
repetitive sequences can prevent aligning reads to unique genomic locations \citep{b506,b519}, and
reads that cannot be uniquely aligned are rejected. In typical
mammalian ChIP-seq experiments, 30 to 40 percent of reads may be discarded, but higher
rates can be encountered in particular experiments. Because of ChIP-seq's cost-effectiveness, such
global losses are usually not an important practical consideration; however, analysis methods
typically make no corrections for the local biases in aligned read densities that are caused by repetitive regions.

Certain types of biases in read density profiles can be estimated by sequencing a 
`control' sample in addition to the immunoprecipitated `treatment' sample, and then using an
analysis method that considers the treatment profile relative to the control profile \citep{b505,b506,b507}.
Considering control data can help identify enriched regions that are false positives, assess
numerical background models, and estimate a threshold for segmenting a read density or `enrichment' profile
in order to identify a subset of significantly enriched regions. Analysis methods can be described
as `two-sample' when a control data set is available and `single sample' when only treatment data are available.

In summary, once reads have been aligned to a reference genome, there are at least four central
analysis issues: interpreting the information in local densities of directional reads;
identifying which high local read densities represent false positives; addressing biases in read
densities that arise from local variations in the efficiency with which reads can be aligned to
unique genomic locations; and segmenting the enrichment profile to identify a statistically and
biologically meaningful subset of enriched regions.

ChIP-seq uses relatively new sequencing technology, and, as was the case while ChIP-chip developed as an experimental approach (e.g. \cite{b54}, \cite{b16}), statistical analysis methods are actively being developed. %\citep{b503,b516}.
\cite{b41} introduced QuEST, a method based on kernel density estimates of the forward and reverse read counts, which allows estimating the length of DNA fragments. The separate forward/reverse profiles are then combined to provide an estimate of binding site locations and to quantify the enrichment. When control sample data are available, QuEST can also estimate a false discovery rate (FDR). Like QuEST, MACS \citep{b42} uses both forward and reverse read profiles to empirically model the `shift size' of ChIP-seq reads, and uses it to improve the spatial accuracy of the predicted binding sites. Instead of using kernel density estimates, MACS uses a parametric model based on a dynamic Poisson distribution to identify and quantify binding events. \cite{b50} introduced a `CisGenome' analysis pipeline for the analysis of ChIP-chip and ChIP-seq data. Their method is also based on a Poisson background model, but includes functionality not offered by MACS and QuEST, e.g. filtering atypical regions, and different types of FDR estimates.

While these methods have established statistical approaches for ChIP-seq analysis, model-based and Bayesian approaches are in earlier stages of development. In the work described here, we introduce a method for probabilistic inference of ChIP-seq data (PICS) that is based on a Bayesian hierarchical truncated $t$-mixture model. PICS integrates four important components. First, it jointly models local concentrations of directional reads. Second, it uses mixture models to distinguish closely-spaced adjacent binding events. Third, it incorporates prior information for the length distribution of immunoprecipitated DNA to help resolve closely adjacent binding events, and identifies enriched regions that have atypical fragment lengths. Fourth, it uses pre-calculated whole-genome read `mappability' profiles to adjust local read densities for reads that are missing due to genome repetitiveness. For each binding event, PICS returns an enrichment score that is relative to a control sample when such a sample is available, and it can use a control sample to estimate a false discovery rate. 
Finally, because it is based on a probabilistic model, PICS can compute measures of uncertainty for binding site estimates, and these can be used to estimate binding site locations and to filter low-confidence regions.

The paper is organized as follows. In section 2, we introduce the data structure and some notation. In section 3, we present our Bayesian hierarchical truncated $t$-mixture model and show how we use it to detect binding events, and to estimate binding site positions
and their confidence intervals. In section 4, we apply PICS to two published, experimental, human ChIP-seq datasets, and compare its results to results from three other methods: QuEST, MACS and CisGenome. In section 5, we briefly discuss our results and possible extensions.
\section{Data, Preprocessing, and Notations}\label{s:data}
We used two ChIP-seq data sets that have been analyzed by other groups. \cite{b42},
using `MACS', identified binding sites of FOXA$1$ (hepatocyte nuclear factor $3\alpha$) in human
MCF$7$ (breast cancer) cells. \cite{b41}, using `QuEST', identified binding sites of the growth
 associated binding protein (GABP) in human Jurkat T cells. Each data set consists of single-end reads for a
 treatment (ChIP) and a control sample. The FOXA1 data consist of $3,909,507$ treatment reads 
 and $5,233,322$ input DNA control reads, while the GABP data consist of $7,830,602$
treatment reads and $17,028,066$ control reads.

Because most of the genome should not interact specifically with a given transcription factor, ChIP-seq aligned-read
data are usually sparse, consisting largely of regions in which few or no reads are observed.
Given this, we first pre-process the read data by segmenting the genome into candidate
regions, each of which has a minimum number of reads that aligned to forward and reverse
strands. We detect such regions using a $w=100$ bp sliding window with an $s=10$ bp step size,
counting the number of forward strand reads in the left half and the number of reverse strand reads in the right half.
Beginning at the start of each chromosome, we retain windows that contain at least one forward
read and one reverse read. For each chromosome, after merging overlapping windows and
removing merged regions with less than two forward or reverse reads, we obtain a disjoint set
of candidate regions, each of which we analyze separately. For the work described here, because
DNA fragments are often between 100 and 300 bp long after gel size selection, we chose $w=100$ bp, and
we set $s=10$ bp for computational convenience. We tested other values for $w$ and $s$ and obtained
essentially the same candidate regions.
\section{Model, priors and parameter estimation}\label{s:model}
In this section, we use ${\cal IG}a(\alpha, \beta)$ to denote an inverse gamma distribution,
and ${\cal G}a(\alpha, \beta)$ to denote a gamma distribution with shape parameter $\alpha$ and
an inverse scale parameter $\beta$. Similarly, $\mathrm{N}(\mu, \sigma^2)$ denotes a Normal
distribution with mean $\mu$ and variance $\sigma^2$, while $t_4 (\mu, \sigma^2)$ denotes
a $t$ distribution with $4$ degrees of freedom, mean $\mu$ and variance parameter $\sigma^2$.

\subsection{Modeling a single binding event}
Having segmented the read data into candidate regions, as described in section \ref{s:data}, we now assume
that each region contains a single transcription factor binding site. An extension to the case of multiple binding sites is
treated below. Let us denote by $f_i$ and $r_j$ the $i-th$ and $j-th$ forward and reverse reads in a given
region, with $i=1,\dots, n_f$  and $j=1,\dots,n_r$. Note that the number of forward reads, $n_f$, and
reverse reads, $n_r$, will typically vary between candidate regions. We jointly model the forward and reverse reads as:
\begin{equation}
  \label{equ:PICSmodel1}
f_i \sim t_4 \left(\mu-\delta/2,\sigma_f^2\right) \quad \mathrm{and} \quad r_j \sim t_4 \left(\mu+\delta/2,\sigma_r^2\right)
\end{equation}
where $\mu$ represents the binding site position, $\delta$ is the
distance between the maxima of the forward and reverse distributions,
which corresponds to the average DNA fragment size of the binding event in question, and
$\sigma_{f}$ and $\sigma_{r}$ measure the corresponding
variability in DNA fragment lengths. Note that this approach differs from that typical for sequencing data,
in that we do not model the sequence counts, but rather the distributions of the fragment ends, for
which we have more prior information. Figure~\ref{f:illu}a displays a candidate region with one binding event,
along with the corresponding PICS parameter estimates.

\subsection{Modeling multiple binding events}
We use mixture models to address the possibility that the sets of forward and reverse reads in single
candidate region were generated by multiple closely-spaced binding events. We model the forward
and reverse reads using $t$-mixture distributions:
\begin{eqnarray}
  f_{i}&\sim&\sum_{k=1}^Kw_k t_4\left(\mu_{fk},\sigma_{fk}^2 \right) \eqd g_f(f_{i}| \boldsymbol{w},\boldsymbol{\mu},\boldsymbol{\delta},\boldsymbol{\sigma}_{f})\nonumber\\
  \label{equ:PICSmodel2} r_{j}&\sim&\sum_{k=1}^Kw_k t_4\left(\mu_{rk},\sigma_{rk}^2 \right) \eqd g_r(r_{j}| \boldsymbol{w},\boldsymbol{\mu},\boldsymbol{\delta},\boldsymbol{\sigma}_{r})
\end{eqnarray}
where $\mu_{fk}=\mu_k-\delta_k/2$ and $\mu_{rk}=\mu_k+\delta_k/2$
and $\mu_{k}$, $\delta_{k}$, $\sigma_{fk}$, $\sigma_{rk}$ are
defined as in (\ref{equ:PICSmodel1}), but have an index $k$
that corresponds to the binding event $k$, while $w_k$ is the mixture
weight of component $k$, which represents the relative proportion of
reads coming from the binding event $k$. For simplicity we denote by
$g_f$ and $g_r$ the resulting p.d.f. of the forward and reverse
mixture distributions.

\begin{figure}
 \centerline{\includegraphics[scale=0.5]{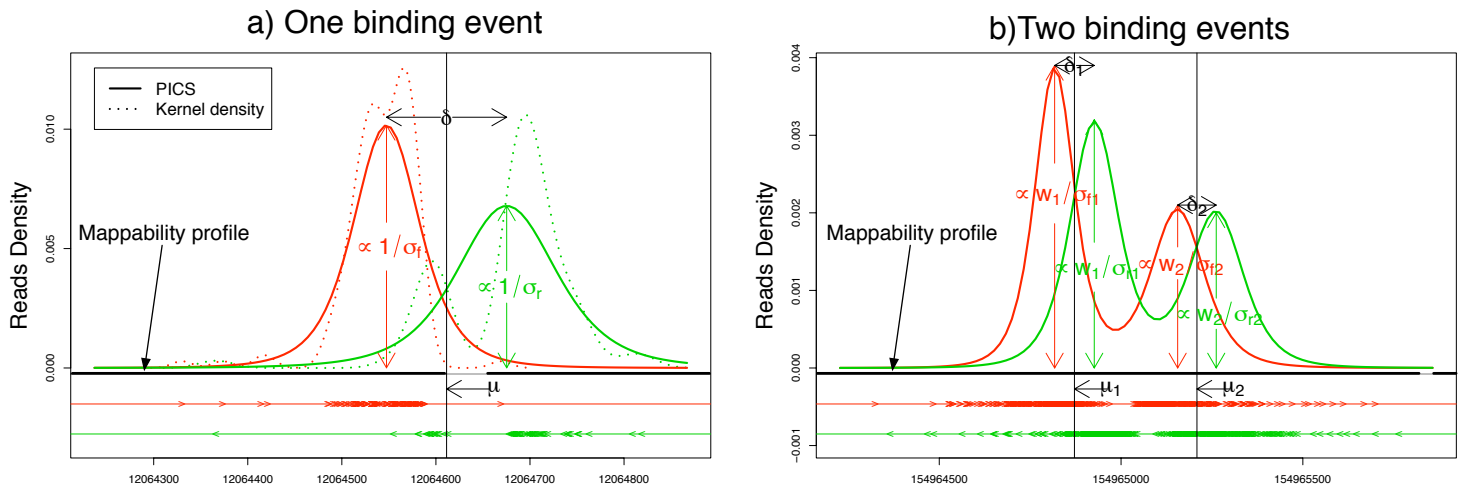}}
\caption{Binding events in two candidate regions in GABP data. PICS detected one binding event in the region in (a) and two binding events in the region in (b). Forward and reverse strand aligned reads are shown by red and green arrowheads, respectively. Mappability profiles are shown as black/white lines, in which the white intervals show nonmappable regions. In (a) the distribution of reverse reads has been biased by a region with low mappability. }
\label{f:illu}
\end{figure}

Figure~\ref{f:illu}b displays a candidate region that has two binding events, along with the corresponding PICS parameter estimates.

% Figure 1

As described in (\ref{equ:PICSmodel1}-\ref{equ:PICSmodel2}),
PICS uses $t$ distributions with 4 degrees of freedom to model local distributions of forward and
reverse reads. While the $t$ distribution is similar in shape to the Gaussian distribution, its heavier
tails make it a robust alternative \citep{b43}. The degrees of freedom are fixed as $v=4$ to minimize
computation \citep{b49}. Note also that since a DNA fragment should contribute a forward read or a
reverse read with equal probability, we use the same mixture weight $w_k$ for both forward and
reverse distributions. Finally, to accomodate possible biases (e.g. in DNA sonication) that result in
asymmetric forward and reverse peaks, we use different forward and reverse variance
parameters $\sigma_{fk}^2$ and $\sigma_{rk}^2$.

% Section 3.3
\subsection{Modeling multiple binding events with missing reads}
\label{s:multiple binding with missing reads}
Building on (\ref{equ:PICSmodel1}-\ref{equ:PICSmodel2}), we now consider the case where some reads are missing due to one or more non-mappable regions intersecting a candidate region. Once again, for illustration, we focus on a single candidate region, whose range is denoted by $S$. For each chromosome, a mappability profile for a specific read length consists of a vector of zeros and ones that gives an estimated read mappability `score' for each base pair in the chromosome \citep{b519}. A score of one at a position means that we should be able to align a read of that length uniquely at that position, while a score of zero indicates that no read of that length should be uniquely alignable at that position. As noted, reads that cannot be mapped to unique genomic locations are typically discarded. For convenience, and because transitions between mappable and non-mappable regions are typically much shorter than these regions, we compactly summarize each chromosome's mappability profile as a disjoint union of non-mappable intervals that specify only zero-valued profile regions (Figure~\ref{f:illu}).

Let us assume that a candidate region intersects one or more of these intervals. We can
write $S=\bigcup_{l=0}^L S_l$, where $S_l=[a_l,b_l]$ denotes
the $l-th$ non-mappable interval, with $l=1,\dots, L$, and $S_0$
denotes the union of intervals that have high mappability, and so should have no
missing reads. In $S_l$, the $f_{li}\; (i=1,\ldots,n_{fl})$ and
$r_{lj}\; (j=1,\dots, n_{rl})$ denote $n_{fl}$ independent forward
reads and $n_{rl}$ independent reverse reads. Note that only the
quantities with $l=0$ are observed, while all others are unobserved
random variables. Also, note that $n_{f0}$, $n_{r0}$, and $L$ will vary across candidate regions.

Based on (\ref{equ:PICSmodel2}), $f_{li}$ and $r_{lj}$, $l=1,\ldots,L$, follow a truncated t-mixture model, which is given by
$g_f$ and $g_r$ truncated on $S_l$. The only information carried in the mappability profile is the location and length of $S_l$; these
affect the estimation of the model parameters shared between the observed and unobserved reads, i.e. $\boldsymbol{w}$,
$\boldsymbol{\mu}$, $\boldsymbol{\delta}$, $\boldsymbol{\sigma}_{f}$, and $\boldsymbol{\sigma}_{r}$. As we will see in Section \ref{s:inf}, it is
possible to account for the missing reads when estimating the unknown parameters.

% Section 3.4
\subsection{Prior distributions}
Typically the library construction process makes prior information  available for the length of the DNA
fragments, $\delta_k$. We can use a Bayesian approach to take advantage of this
information by allowing the $\delta_k$'s for all binding sites to derive from a common prior
fragment length distribution.
Similarly, we can also put a common prior distribution on $\sigma_{fk}^2$ and $\sigma_{rk}^2$, which
allows us to incorporate prior information about the variability of the DNA fragment length within a
site and to regularize variance estimates when few reads are available. For computational convenience,
we use a normal inverse gamma conjugate prior, given by
 \begin{equation}
   \label{equ:PICSprior}
 \sigma_{fk}^2,\; \sigma_{rk}^2  \sim {\cal IG}a(\alpha,\beta) \quad \mathrm{and} \quad (\delta_k|\sigma_{fk}^2,\; \sigma_{rk}^2)\sim\mathrm{N}(\xi,\rho^{-1}/(\sigma_{fk}^{-2}+\sigma_{rk}^{-2}))
 \end{equation}
where $\xi$ represents our best prior guess about the mean fragment length across all binding sites, and $\rho$ controls the spread around this guess. Similarly, $\beta/(\alpha-1)$ represents our best prior guess about the variance of the DNA fragment length, and $\beta^2/(\alpha-1)^2(\alpha-2)$ controls the spread around this prior guess. In the analysis reported here, we choose $\alpha=20$, $\beta=40000$, $\xi=175$, and $\rho=1$. These values were based on knowing that DNA fragments should be on the order of 100-250 bps after gel size selection for both datasets considered \citep{b41,b42}, and resulted in a fairly non-informative prior for the DNA fragment length, with a mean of 175 bps and a standard deviation of approximately 50 bps.

\subsection{Parameter Estimation Using the EM Algorithm}
Given the conjugacy of the prior chosen, an Expectation-Maximization
(EM) algorithm can be derived to find the maximum likelihood estimates (MLE) of the unknown parameter vector
$\boldsymbol{\bTheta}=(\btheta_1,\dots,\btheta_K)$ where $\btheta_k=(
w_k, \mu_k, \delta_k, \sigma_{fk}^2,, \sigma_{rk}^2)$. Our algorithm is similar to those used in $t$ mixture models
and Bayesian regularization for mixture models  \citep{b46, b44}. In the presence of missing reads, we use an
algorithm similar to that developed by \cite{b45} for grouped and truncated data. In the following text, for ease of
notation, we use the letter $d$ to denote either $f$ or $r$. For simplicity, we first describe our EM algorithm when
no missing reads are present, i.e. for $S=S_0$, $d_{li}=d_i$.

\noindent\textbf{Complete data likelihood:} We consider the
`complete data' to be $\by_{di}=\left(d_{i}, \bz_{di},
\bu_{fi}\right)$, where $\bz_{di}$ and $\bu_{di}$ are the missing
data. The newly introduced missing data are: first, the unobserved cluster
memberships, which are defined as $\bz_{di}=\left(z_{di1}, \dots,
z_{diK}\right)$ for the reads, where $z_{dik}$ is a binary indicator
that the read $d_{i}$ belongs to mixture component $k$;
and second, the weights $\bu_{di}=(u_{di1}, \dots, u_{diK})$, which come from the normal-gamma compound parameterization, and are defined by
\begin{eqnarray}
(d_{i}|U_{dik}=u_{dik},z_{dik}=1,\mu_k,\delta_k) &\sim& \mathrm N\left(\mu_{dk},\frac{\sigma_{dk}^2}{u_{dik}}\right)\label{eqn:model-f} \\
(U_{dik}| z_{dik}=1) &\sim& {\cal G}a(v/2,v/2), \label{eqn:model-r}
\end{eqnarray}
independently for $i=1,\dots,n_{di}$, where $v=4$ is the degrees of freedom of the $t$ distribution. The advantage of writing the model in this way is that, conditional upon the $\bu_{di}$'s, the sampling errors are again normal but with different precisions, and estimation becomes a weighted least squares problem.

The penalized log complete data likelihood, denoted $l^*$, is given by
% \begin{equation}\label{eqn:comp-loglik}
$
      l^*(\bTheta|\by)=l(\bTheta|\by)+l_{\mbox{prior}},
$
% \end{equation}
where $l(\bTheta|\by)$ is the complete-data log-likelihood, given by
\begin{eqnarray*}\label{equ:CompleteLikelihood}
&&l(\bTheta|\by)\\
&=&\sum_{d \in \{f,r\}} \sum_{i=1}^{n_{d}}\sum_{k=1}^G z_{dik}\left\{\log\left[w_k\mathrm{N}\left(d_{i}|\mu_{dk},\frac{\sigma_{dk}^2}{u_{dik}}\right){\cal G}a(u_{dik}|2,2)\right]\right\}\\
&=&\sum_{d \in \{f,r\}} \sum_{i=1}^{n_{d}}\sum_{k=1}^G z_{dik}\left\{\log w_k - \log\sigma_{dk} - \log \sqrt{2\pi} -\frac{u_{dik}}{2} \left(\frac{d_{i}-\mu_{dk}}{\sigma_{dk}}\right)^2+ \log u_{dik}-2u_{dik}+\log4\right\},
\end{eqnarray*}
and $l_{\mbox{prior}}$, the log prior `penalty' on $(\boldsymbol{\delta},\boldsymbol{\sigma}_{f}^2,\boldsymbol{\sigma}_{r}^2)$, is given
as
\begin{equation}
l_{\mbox{prior}}=-\frac{1}{2}\sum_k\left\{(\sigma_{fk}^{-2}+\sigma_{rk}^{-2})[\rho
(\delta_k-\xi)^2+2\beta]\right\}+\frac{2\alpha-1}{2}\sum_k\left\{\log(\sigma_{fk}^{-2}+\sigma_{rk}^{-2})\right\}.
\end{equation}

\noindent\textbf{E-Step:} Given the current estimate $\bTheta^-$ for $\bTheta$, the conditional expectation of the penalized log complete data likelihood is given as
\begin{eqnarray}
\nonumber   Q(\bTheta|\bTheta^-) &\eqd& \mathbb{E}[l(\bTheta|\by)|\bTheta^-]+l_{\mbox{prior}}\\
\label{eqn:Ql}          &=& \sum_{d \in \{f,r\}} \sum_{i=1}^{n_{d}}\sum_{k=1}^K \tilde{z}_{dik}\left\{\log w_k - \log\sigma_{dk} -\frac{\tilde{u}_{dik}}{2} \left(\frac{d_{i}-\mu_{dk}}{\sigma_{dk}}\right)^2\right\}+A
\end{eqnarray}
%IN THIS CASE I THINK THE CONSTANT DOES NOT DEPEND ON THETA RIGHT?
where $A$ is a constant with respect to the parameter vector $\bTheta$. Given this,
the E-step \citep{b44} consists of computing the following quantities
\begin{eqnarray}
\label{eqn:tildezf}  \tilde{z}_{dik}  &\eqd & \mathbb{E}(Z_{dki} |\by_{di},\bTheta^-)  =\frac{w_k t_{4}(d_{i}|\mu_{dk},\sigma_{dk})}{\sum_k w_k t_{4}(d_i|\mu_{dk},\sigma_{dk})},\\
%\nonumber  &=&  \frac{w_k t_{4}(f_{li}|\mu_{dk},\sigma_{dk})}{\sum_k w_k t_{4}(d_i|\mu_{dk},\sigma_{dk})}, \\
\label{eqn:tildeuf}  \tilde{u}_{dik}  &\eqd &
\mathbb{E}(U_{dik}|\by_{di},z_{dik}=1,\bTheta^-)
=\frac{5}{4+(d_{i}-\mu_{dk})^2/\sigma_{dk}^{2}}.
%\nonumber  &=& 5/[4+(d_{li}-\mu_{dk})^2/\sigma_{fk}^{2}]. \\
\end{eqnarray}

\noindent\textbf{M-Step:} During the M-step, the goal is to maximize $Q(\bTheta|\bTheta^-)$ with respect to $\bTheta$, which requires solving
% \begin{equation}\label{m-step}
$
 \partial Q(\bTheta|\bTheta^-)/\partial \bTheta=\boldsymbol{0}.
$
% \end{equation}

Unfortunately, there is no simple closed form solution for $\bTheta$. Given this, we adopted a conditional approach in which we first maximize over $(\mathbf{w},\bmu,\bdelta)$, conditional on $(\bsigma_{f}, \bsigma_{r})$, and then maximize over $(\bsigma_{f}, \bsigma_{r})$, conditional on the previously updated $(\mathbf{w},\bmu,\bdelta)$, resulting in an Expectation/Conditional Maximization (ECM) algorithm \citep{Rubin}. Conditional on $\sigma_{fk}$ and
$\sigma_{rk}$, we solve a linear system analytically, which leads to the following estimates:
\begin{eqnarray*}
  \hat{w}_k &\leftarrow& \frac{\tilde{\chi}_{fk}+\tilde{\chi}_{rk}}{N_f+N_r}, \\
  \hat{\mu}_k &\leftarrow& \frac{\tilde{s}_{fk}+\tilde{s}_{rk}}{\tilde{m}_{fk}+\tilde{m}_{rk}}+\frac{\tilde{m}_{fk}-\tilde{m}_{rk}}{2(\tilde{m}_{fk}+\tilde{m}_{rk})}\hat{\delta}_k, \\
  \hat{\delta}_k &\leftarrow& \frac{\tilde{s}_{rk} \tilde{m}_{rk}^{-1}-\tilde{s}_{fk}\tilde{m}_{fk}^{-1} + \rho (\hat{\sigma}_{fk}^{-2}+\hat{\sigma}_{rk}^{-2}) \xi(\tilde{m}_{fk}^{-1}+\tilde{m}_{rk}^{-1})}{1+ \rho (\hat{\sigma}_{fk}^{-2}+\hat{\sigma}_{rk}^{-2}) (\tilde{m}_{fk}^{-1}+\tilde{m}_{rk}^{-1})},
\end{eqnarray*}
where
\begin{equation}
\tilde{\chi}_{dk} = \sum_{i=1}^{n_{d}}\tilde{z}_{dik},\;\; \tilde{s}_{dk}=\sum_{i=1}^{n_{d}}d_{i} \tilde{z}_{dik}\tilde{u}_{dik},\;\; \tilde{m}_{dk}=  \sum_{i=1}^{n_{d}}\tilde{z}_{dik}\tilde{u}_{dik}.\nonumber
\end{equation}
Conditional on these new estimates $\hat{w}_k,\hat{\mu}_k,\hat{\delta}_k$, we can then solve a non-linear system analytically. The new estimate of $\sigma_{dk}^{-2}$ is the only non-negative root, and is given as
\[
  \hat{\sigma}_{dk}^{-2} \leftarrow \left(C_{3d}-C_1\right) /C_{4d},
\]
where
\begin{eqnarray}
\tilde{\eta}_{dk}&=&  \sum_{i=1}^{n_{d}}(d_{i}-\hat{\mu}_{dk})^2 \tilde{z}_{dik}\tilde{u}_{dik} \nonumber \\
C_{2d}  &=& \rho (\hat{\delta}_k-\xi)^2+2\beta+\tilde{\eta} _d,\nonumber \\
C_{3f}&=& (\tilde{\eta} _f -\tilde{\eta} _r) \left(2 \alpha -1+\tilde{\chi} _f\right)+ C_{2f} \left(\tilde{\chi} _f+\tilde{\chi} _r\right),\nonumber \\
C_{3r}&=& (\tilde{\eta} _r -\tilde{\eta} _f) \left(2 \alpha -1+\tilde{\chi} _r\right)+ C_{2r} \left(\tilde{\chi} _r+\tilde{\chi} _f\right),\nonumber \\
C_{4d}  &=& 2 C_{2d} \left(\tilde{\eta} _f-\tilde{\eta} _r\right),\nonumber \\
C_1 &=& \sqrt{\left[(2 \alpha-1) (\tilde{\eta} _f-\tilde{\eta} _r) +\nonumber
C_{2f} \tilde{\chi} _r - C_{2r}\tilde{\chi} _f \right]^2+4 C_{2r}\nonumber
\tilde{\chi} _f C_{2f} \tilde{\chi} _r}.
\end{eqnarray}

\noindent\textbf{Accounting for missing reads:} In the presence of missing reads, we decompose the log complete data likelihood as $l(\bTheta|\by)=\sum_{l=0}^L l_l(\bTheta|\by)$,
where $l_l(\bTheta|\by)$ is the complete-data log-likelihood in
partition $S_l$, given by
\begin{eqnarray*}\label{equ:CompleteLikelihoodMiss}
&&l_l(\bTheta|\by)\\
&=&\sum_{d \in \{f,r\}} \sum_{i=1}^{n_{dl}}\sum_{k=1}^G z_{dlik}\left\{\log w_k - \log\sigma_{dk} - \log \sqrt{2\pi} -\frac{u_{dlik}}{2} \left(\frac{d_{li}-\mu_{dk}}{\sigma_{dk}}\right)^2+ \log u_{dlik}-2u_{dlik}+\log4\right\}.
\end{eqnarray*}

We now have additional missing data, $n_{dl}$ and $d_{li}$, corresponding to the number of missing reads and the missing reads themselves.
% In the presence of missing reads, $d_{li}=d_l$ and $n_{dl}$ become unobserved random variable as
% discussed in section~\ref{s:multiple binding with missing reads}.
To accommodate this, all that we need to change is to add two steps to our E-step, as follows.

% First, assuming 
Because the unknown counts $n_{dl}$,  $l=1,\ldots,L$,
follow a negative multinomial distribution, we simply replace them with their conditional expectations, which are given by
\begin{eqnarray}
\label{eqn:tildenf}  \tilde{n}_{dl}    \eqd
\mathbb{E}(n_{dl}|\by_{d0i},\bTheta^-)= n_{d0}
P_{dl}(\bTheta^-)/P_{d0}(\bTheta^-),
\end{eqnarray}
where $P_{dl}(\bTheta^-) \eqd \Pr(X \in S_l) = \int_{S_l}
g_d(x|\bTheta^-) dx$ and $P_{d0}(\bTheta^-)\eqd \Pr(X \in S_0)
=1-\sum_{l=1}^L P_{dl}(\bTheta^-)$ are the probability measures of
the partitions $S_l$ and $S_0$.

Second, conditional on the imputed counts $ \tilde{n}_{dl}$, we replace the following quantities with the corresponding expectations
\begin{eqnarray*}
\tilde{\chi}_{dk}&\leftarrow&\tilde{\chi}_{d0k}+\sum_{l=1}^L\tilde{n}_{dl}\mathbb{E}_{dl}[\tilde{z}_{dlk}],\\
\tilde{s}_{dk}   &\leftarrow&\tilde{s}_{d0k}   +\sum_{l=1}^L\tilde{n}_{dl}\mathbb{E}_{dl}[\tilde{z}_{dlk}\tilde{u}_{dlk}],\\
\tilde{m}_{dk}   &\leftarrow&\tilde{m}_{d0k}   +\sum_{l=1}^L\tilde{n}_{dl}\mathbb{E}_{dl}[d_l\tilde{z}_{dlk}\tilde{u}_{dlk}],\\
\tilde{\eta}_{dk}&\leftarrow&\tilde{\eta}_{d0k}+\sum_{l=1}^L\tilde{n}_{dl}\mathbb{E}_{dl}[(d_l-\hat{\mu}_k)^2\tilde{z}_{dlk}\tilde{u}_{dlk}],
\end{eqnarray*}
where $\tilde{\chi}_{d0k}$, $\tilde{s}_{d0k}$, $\tilde{m}_{d0k}$, and $\tilde{\eta}_{d0k}$ are the original quantities as defined in M-step in the case of no missing reads, and $\mathbb{E}_{dl}$ are the expectations with respect to the unobserved reads ($d_{li}$, $l>0$),
conditional on observed reads $d_{0i}$ and on previous estimated parameters $\bTheta^-$ (the Appendix gives
% , section ~\ref{s:detail extra E-step},
more details of computing these expectations).

\subsection{Inference and Detection of Binding Sites}

\noindent\textbf{Choosing the number of binding events in each region:}
The EM algorithm described above assumes that the number of binding events within a region, $K$, is known. However, in practice, $K$ is unknown and needs to be estimated. For each candidate region, we fit our PICS model
% model (\ref{equ:PICSmodel1})
with $K$ taking values from 1 to 15, and select the value of $K$ that has the largest BIC \citep{b51}, which in our case is given by
\begin{equation}\label{eqn:BIC}
  BIC=-2Q(\bTheta=\hat{\bTheta}|\hat{\bTheta})+(5K-1)\ln(n_{f0}+n_{r0}),
\end{equation}
where $\hat{\bTheta}$ is the final estimate for the parameters
$\bTheta$.

\noindent\textbf{Uncertainty of parameter estimates:}
It is useful to extend the point estimates for the parameters of interest, $\boldsymbol{\mu}$ and $\boldsymbol{\delta}$, by deriving measures of uncertainty for them. Within our framework of mixture models with truncated data, we derive an approximation of the observed
information matrix for the parameters using the approach described in \cite{b48}. Using the observed information matrix, we can then obtain approximate standard errors for both $\boldsymbol{\hat{\mu}}$ and $\boldsymbol{\hat{\delta}}$.
We can use these standard errors to, for example, define the starts and ends of
binding event neighborhoods, filter out
noisy enriched regions and estimate confidence intervals for binding site point locations.

\noindent\textbf{Binding event neighborhoods:}
Because PICS models local concentrations of bidirectional reads, we can define `high confidence' neighborhoods whose extents are given by the maxima of forward and reverse density distributions. Using our PICS parameters, and taking into consideration the standard errors of the estimates, for a given binding event this neighborhood is defined as the interval $\mu\pm\delta/2$, extended by three standard errors on each side, i.e. (SE($\mu-\delta/2$) for the left limit and SE($\mu+\delta/2$) for the right limit).
These high confidence neighborhoods can define `enriched' regions in a file that can be visualized in a genome browser \citep{b520}.

\noindent\textbf{Peak merging and filtering:} We use BIC to estimate the number of binding
events within each candidate region. While BIC is well suited to selecting the number of mixture
components required to estimate an underlying probability density, it can sometimes overestimate
the number of components \citep{b56}. In our case, when a candidate region contains hundreds
of reads, BIC may select a model that has too many components in obtaining a good fit to the underlying density.
To address this, we merge peaks that have overlapping binding events, as defined by the start
and end positions defined above. The parameters of the merged peaks are obtained by moment
matching conditions (see appendix). Since the combined parameters $\bmu$ and $\bdelta$ are
linear combination of the original ones, the original information matrix can be used to recompute
the standard errors. For the GABP and FOXA1 data described below, this approach merged less
than 1\% of the binding events.

In addition to merging overlapping events, we also filter out binding events that have noisy
or atypical parameter estimates, which could potentially affect the downstream analysis. Specifically, we remove
binding events that fail to satisfy any of the following three criteria:
\[
  (i)\hspace{2mm}\mbox{SE}(\mu) < 50;\;\;\;  (ii)\hspace{2mm}50 < \delta < 200;\;\;\; (iii)\hspace{2mm}\sigma_f, \sigma_r <
  150.
\]
Essentially, $(i)$ filters events that have noisy binding site position estimates, $(ii)$ filters events with atypical average DNA fragment length estimates (e.g. events that have high fractional overlaps with simple tandem repeats \citep{b530}), and $(iii)$ filters events with large DNA fragment length variability.

\noindent\textbf{Scoring and ranking binding events:}
In order to identify and rank a statistically meaningful subset of binding events, we define an enrichment score for each binding event. For a given event, we define $F_{ChIP}$ and $R_{ChIP}$, the number of observed forward/reverse ChIP (`treatment') reads that fall within the 90\% contours of the forward/reverse distributions, i.e. within $\mu_d \pm 2.13 \sigma_d$. We assign an enrichment score to each binding event as $s=\min(F_{ChIP},R_{ChIP})$. When a control sample is available, we similarly define $F_{control}$ and $R_{control}$, by computing the number of observed forward/reverse reads in the control sample that fall within the 90\% contour of the forward/reverse distributions estimated from the ChIP sample. Using this information, we define an enrichment score for the treatment relative to the control as $s=(N_{control}/N_{ChIP}) \cdot \min_{\{d=F,R\}}((d_{ChIP}+1)/((d_{control}+1)))$, where the addition of the constant one prevents a division by zero. The scaling of the enrichment score by $N_{control}/N_{ChIP}$ accounts for the control and ChIP samples having different numbers of reads.

\noindent\textbf{False discovery rate:} Given control data, we can estimate the false discovery rate as a function of the enrichment score.
We do this by simply repeating the analysis after swapping the control sample for the ChIP sample
and recomputing our enrichment scores, which we call `null' enrichment scores and denote by $s_0$. Then the FDR, as a function of the threshold value $q$, can be computed as follows:
\[
FDR(q)=\frac{\{\#s_0: s_0>q\}}{\{\#s: s>q\}}.
\]

\section{Application to experimental datasets}\label{s:inf}
We applied PICS to the two experimental data sets described in section \ref{s:data}, obtaining 58,622 candidate regions and 60,087 binding events for GABP data, and 32,287 candidate regions and 32,418 binding events for FOXA1 data. Table \ref{t:MultipleMotif} summarizes the number of binding events, broken down by the number of mixture components detected in the corresponding candidate region. Most of the candidate regions were estimated to contain a single binding event, but a non negligible number may contain more than one event. For example, for GABP, 2274 of the binding events that PICS detected were in two-event candidate regions. 
The table also suggests that PICS' mixture model was effective in discriminating closely-spaced binding events, as, for example, for the top-ranked 5000 GABP events, 79 percent of events in two-component regions were associated with a predicted GABP motif site (see below for details about motif sites).

\begin{table}
\caption{Number of PICS binding events found for the GABP and FOXA1 data,
broken down by the number of mixture components detected in the corresponding
candidate region. The first two rows give, for the 5000 most
significant binding events for each data set, the number of events identified in regions that had 
1, 2 or 3+ mixture components, and, for each of these classes of events, the percentage of events 
that was associated with at least one predicted site motif site. For example, in the 5000 
top-ranked GABP regions, of the 903 binding events in two-component regions, 79 percent could 
be associated with a predicted GABP site. }
\label{t:MultipleMotif}
\begin{center}
\begin{tabular}{lccccccc}
\hline\hline
 & \multicolumn{3}{c}{GABP} && \multicolumn{3}{c}{FOXA1}\\
\# of components in region & 1 & 2 & 3$+$ && 1 & 2 & 3$+$ \\\hline
\# of events (top 5000 regions) & 3829 & 903 & 64 && 4913 & 74 & 3\\
\% of motifs (top 5000 regions) & 77 & 79 & 73 && 81 & 75 & 67\\
\# of events (all regions)  & 56,229 & 2274 & 119 && 32,012 & 266 & 9\\
\hline
\end{tabular}
\end{center}
\end{table}

\begin{figure}
 \includegraphics[scale=0.5,angle=270]{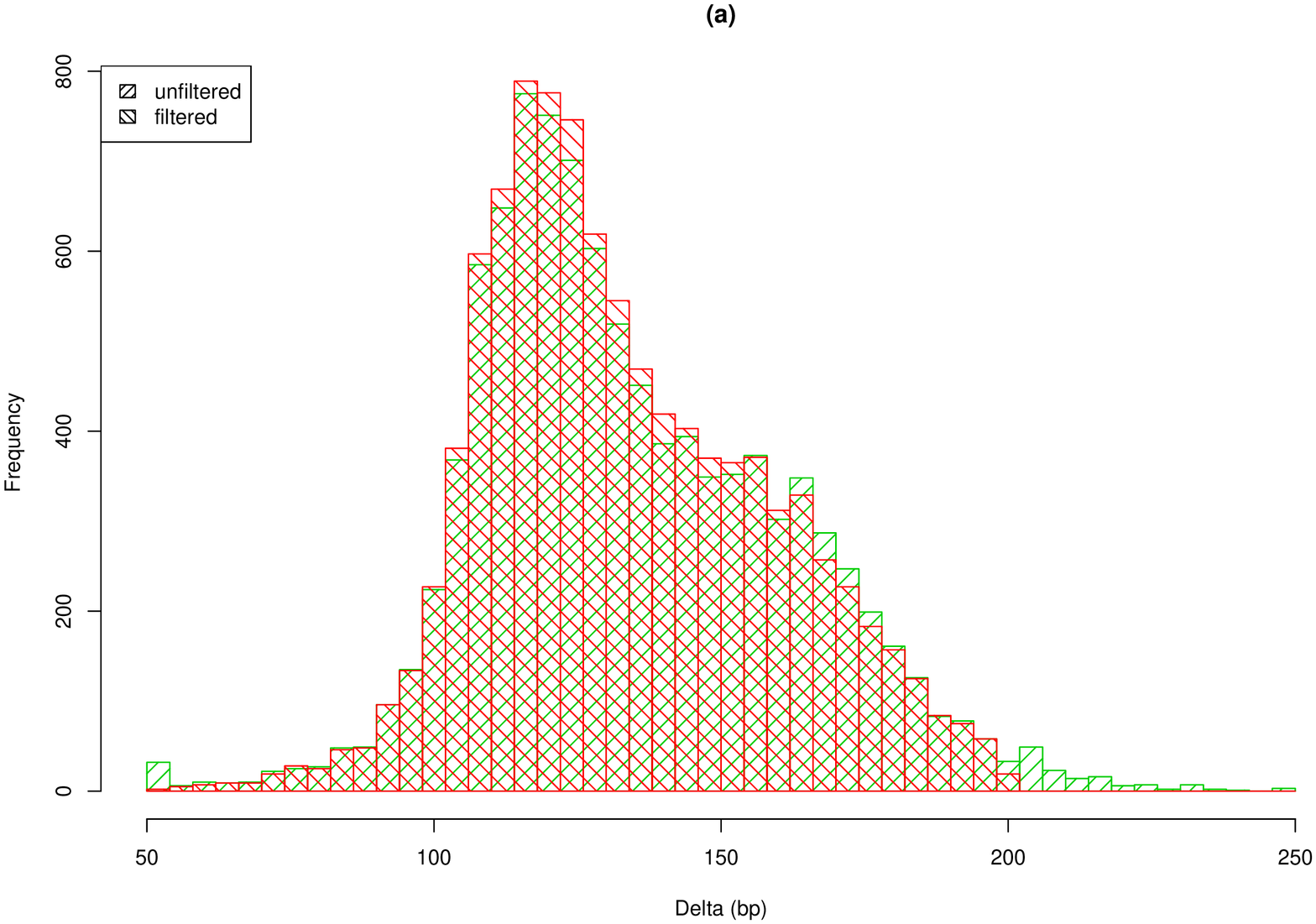}
 \includegraphics[scale=0.5,angle=270]{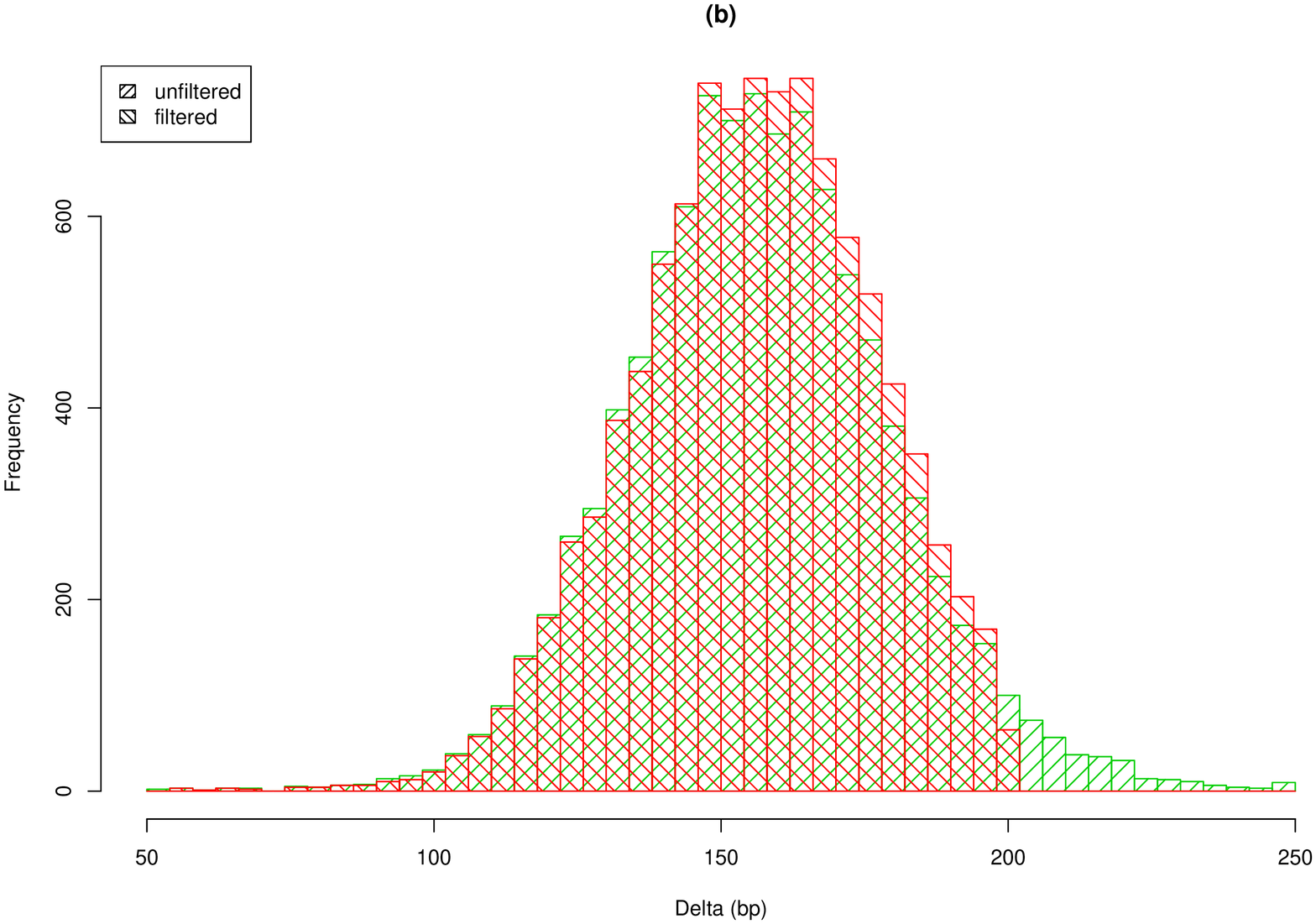}
\caption{Histogram of estimated average DNA fragment lengths,  $\delta$, in GABP (a) and FOXA1 (b) data, before and after filtering. For clarity, only results for the top 10000 regions are shown.} \label{f:hist}
\end{figure}

\begin{figure}
 \centerline{\includegraphics[scale=0.6,angle=270]{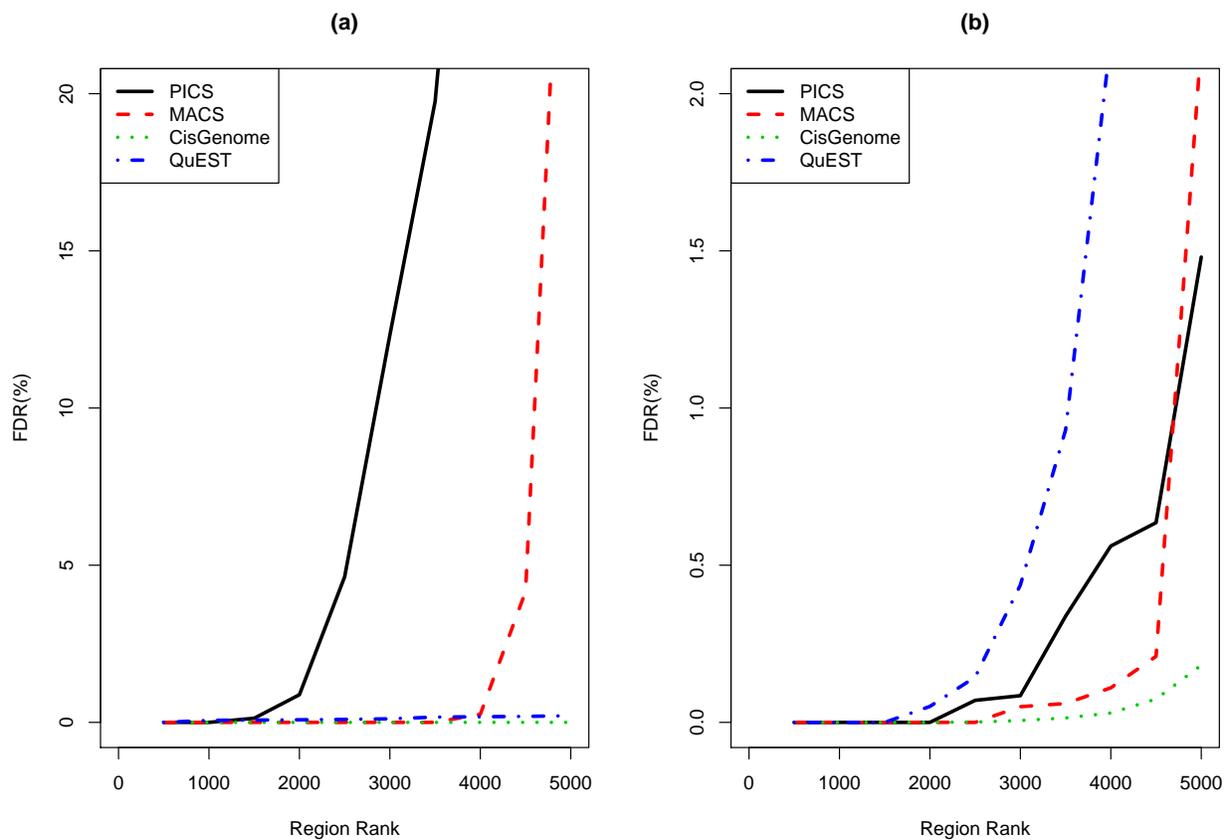}}
\caption{Number of detected peaks at different False Discovery Rate levels for the four analysis methods, for GABP data (a) and FOXA1 data (b).}
\label{f:FDR}
\end{figure}

Figure~\ref{f:hist} shows histograms of estimated average DNA fragment
lengths $\delta$ for the top-ranked 10000 filtered and unfiltered enriched regions.
We considered only this subset, because, based on the estimated FDR (Figure \ref{f:FDR}), the
other regions are likely to be false positives. For the FOXA1 data the estimated average fragment size
 was approximately 150 bps, consistent with \cite{b42};  it was somewhat smaller for the GABP data. 
 Figure~\ref{f:hist} also shows that most of the regions had
DNA fragments between 50 and 200 bps, which supports our filtering atypical regions by this parameter.

We now compare the performance of PICS and the QuEST, MACS and 
CisGenome analysis methods, using the FOXA1 data and GABP data.
Figure~\ref{f:FDR} shows the relationships between the region rank and FDR
for the top-ranked 5000 regions for each method. As expected, the top-ranked regions 
for all methods had FDRs whose values were very small or zero. While CisGenome was 
consistent in returning the largest number of low-FDR regions for both datasets, the responses of
the other three methods differed for GABP and FOXA1 data. QuEST's response
was markedly different for the two sets of data, being close to
CisGenome's for GABP, but having the smallest number of low-FDR regions for
FOXA1. MACS' response was similar to those of QuEST and CisGenome for the
first 4000 GABP regions, after which its response was approximately parallel to 
that of PICS. For FOXA1, the MACS curve diverged progressively from CisGenome's after approximately 2500 regions, then changed slope abruptly at approximately 4500 regions and crossed PICS'
curve. PICS returned by far the fewest low-FDR regions for GABP data, but
its response to FOXA1 data was intermediate between that of QuEST and MACS
for ranks between 2000 and approximately 4500.

Noting that the algorithms could respond very differently to different data
sets in terms of FDR, we then compared the four methods by identifying conserved DNA sequence motifs
in the 5000 top-ranked predictions from each method, using $200$-bp wide regions that
were centered on each method's binding site estimates (`peak summits'). For motif analysis we used GADEM  \citep{b55}, which
can process large sets of ChIP-seq regions on a single CPU, identifies multiple motifs and adjusts
motif widths, and performs well relative to algorithms that are more computationally
demanding. We assessed the \emph{de novo} motifs using STAMP \citep{b501},
and retained only `expected' and biologically relevant motifs.
As expected, for all four methods, GADEM identified GABP and Forkhead motifs as
the dominant motifs in GABP and FOXA1 datasets respectively. For the FOXA1 data, 
regions for all methods also contained the binding motif for the FOS proto-oncogene protein. 
The FOS gene family encodes leucine zipper proteins that can dimerize with proteins of 
the JUN family to form the AP-1 complex \citep{Milde}. The AP-1 complex is over-expressed 
in ER positive cells (\emph{e.g.} MCF7) and can interact directly with the ER transcription factor \citep{Milde,Cicatiello}. Similarly, the FOXA1 protein is known to play an important role in ER regulation and to interact with ER \citep{eeckhoute,lupien}. The FOS motif that we identified was consistent with AP-1 enriched motifs reported for ChIP-chip FOXA1 regions \cite{lupien} and may reflect interactions, possibly indirect, between the FOS and FOXA1 proteins. All other motifs identified by GADEM appeared to be due to repetitive elements.
For the work described here, we used GABP motif occurrences for evaluating GABP results, and
both FOX and FOS motif occurrences for evaluating FOXA1 results.

We evaluated the four methods using two criteria: 1) the motif occurrence rate, i.e. the fraction of
enriched regions that contained a biologically `expected' motif, for which a larger value indicates better performance;
and 2) the spatial accuracy, i.e. the distance between a binding site point estimate and a motif occurrence,
for which a smaller value indicates better performance. Because a motif can occur more than once
in a sequence, we used only the motif instance closest to the predicted
binding event (peak summit) when computing the spatial accuracy.

\begin{figure}
 \centerline{\includegraphics[scale=0.7,angle=270]{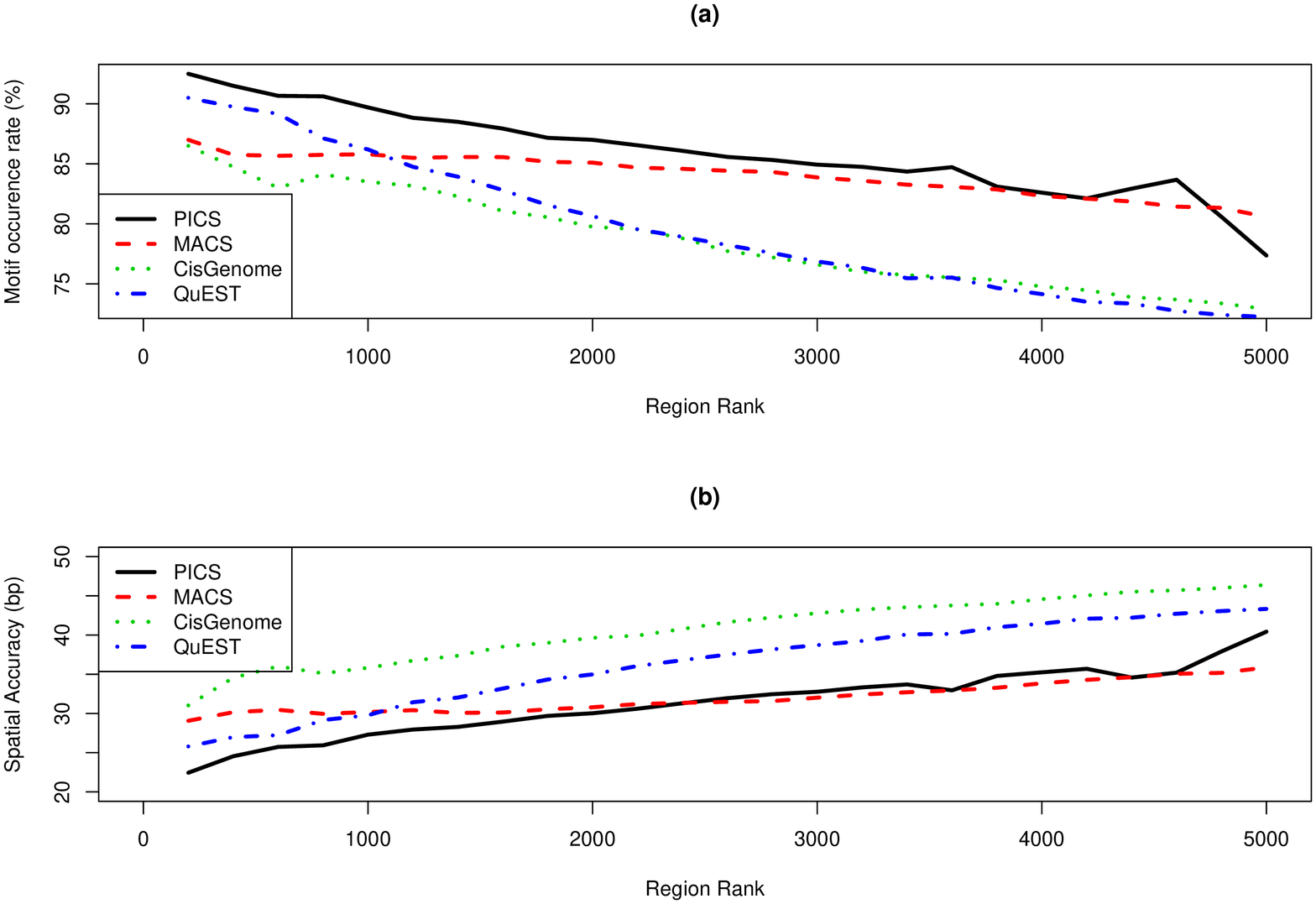}}
\caption{Motif occurrence rate and spatial accuracy for GABP data, as a function of
enriched region rank, for the 5000 top-ranked regions for each method.} \label{f:quest}
\end{figure}

Figures~\ref{f:quest}a,b show the motif occurrence rate and spatial accuracy as a
function of the region rank, for each methods' top-ranked 5000 enriched GABP regions.
PICS had the highest motif occurrence rate for ranks above approximately 3500, below 
which PICS' and MACS' rates appeared comparable. MACS' rates were intermediate for 
ranks between 1000 and 3800, but below QuEST's rate for ranks above 1000. Rates 
for QuEST and CisGenome were lower, and were comparable for ranks below 2000. PICS 
and MACS had the best spatial accuracy, with PICS more accurate for ranks above
 2000, followed by QuEST and CisGenome.

% Figure 4

\begin{figure}
 \centerline{\includegraphics[scale=0.7,angle=270]{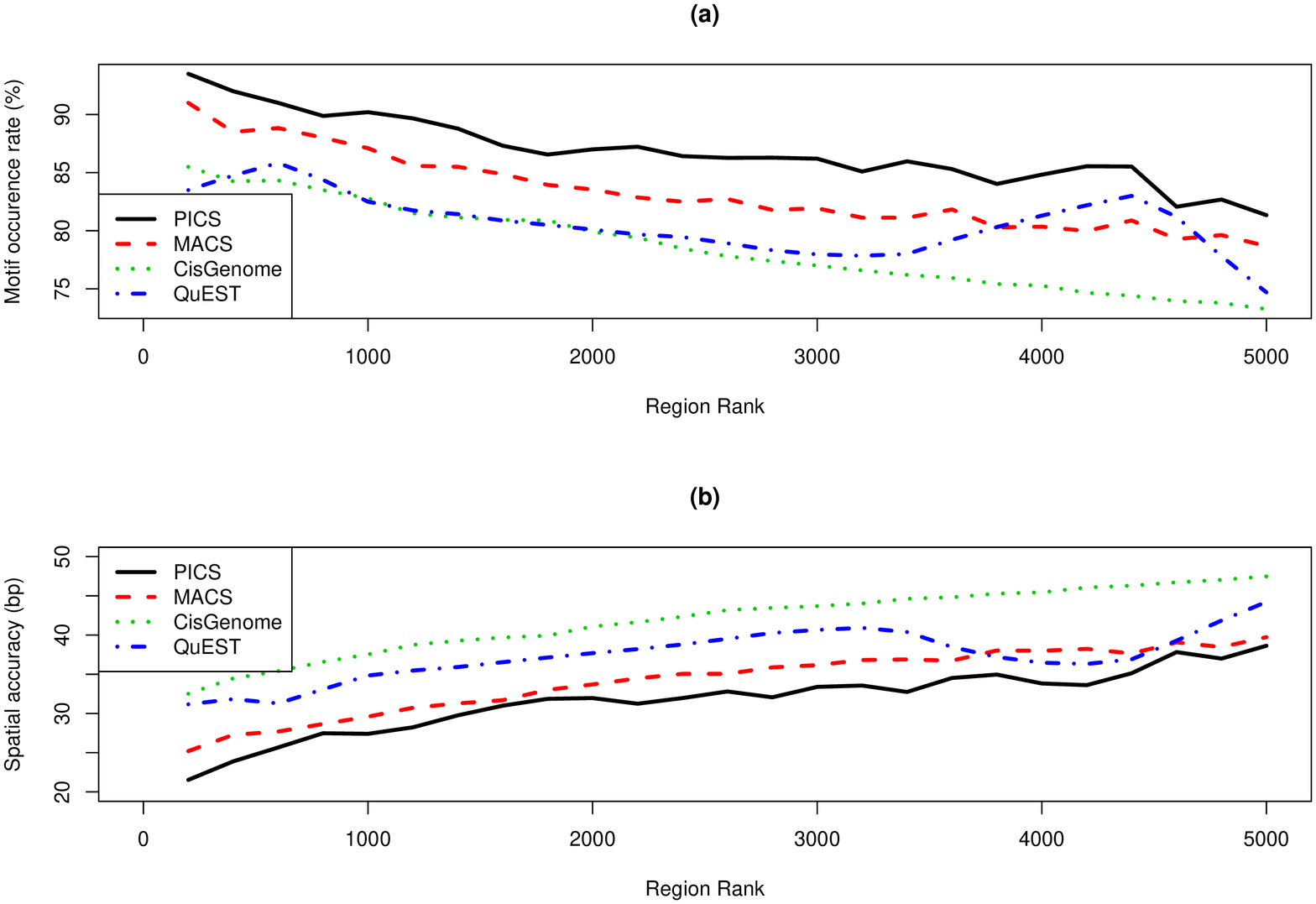}}
\caption{Motif occurrence rate and spatial accuracy for FOXA1 data, as a function of
enriched region rank, for the 5000 top-ranked regions for each method.} \label{f:foxa1}
\end{figure}

% Figure 4
Figures~\ref{f:foxa1}a,b show motif occurrence rate and spatial accuracy for the
FOXA1 data. Considering both metrics over the full range of the top 5000 regions, the
relative performance of the four methods was generally similar to that for GABP data: 
PICS, followed by MACS, QuEST and then CisGenome. 

% Figure 5

Because cells can use multiple closely-spaced transcription factor binding sites
to establish progressive expression responses to cellular signals,
we assessed how effectively PICS' mixture model can detect closely
adjacent binding sites.
Using our predicted transcription factor binding motifs for the top-ranked 5000 PICS
predictions for GABP and FOXA1 data, we determined the
percentage of binding events from single- and multiple-component candidate
regions that could be associated with at least one motif site. Table
\ref{t:MultipleMotif} shows these results as a function of the number of mixture
components in a region. Far more GABP regions than FOXA1 regions had two components (903 vs. 74)
or at least three components (64 vs. 3). For both data sets, the percentage of binding events that was associated with a
predicted binding motif was relatively insensitive to the number of mixture components in a region.
These results suggest that our mixture model was effective in distinguishing
biologically meaningful proximal binding events.

To assess the ability of the other methods to detect proximal binding events we generated
a similar table, but this time considered binding events that had at least one other event within
a fixed distance $d$. Table \ref{t:ProximalEvents} summarizes the
results for $d=250, 500$ and $1000$ bps. For these data, PICS and QuEST were the most effective
at identifying proximal binding events, and a large fraction of these events was
associated with a predicted motif site. While QuEST predicted the largest number of proximal
binding sites, a larger fraction of the mixture components reported by PICS were associated with predicted
binding motifs. For these data, MACS and CisGenome were less effective at discriminating closely spaced binding events.

\begin{table}
\caption{Number of proximal binding events found by in the 5000 top-ranked regions identified by each method
in GABP and FOXA1 data, as a function of the motif `proximity'
distance $d$. The numbers in paratheses give the percentage
of binding events that could be associated with at least one predicted motif site.
For example, the first row ($d=250$) gives the number and percentage of events that had at least
one other binding event within 250 $bp$.}
\label{t:ProximalEvents}
\begin{center}
\begin{tabular}{lccccccccc}
\hline\hline
 & \multicolumn{4}{c}{GABP} && \multicolumn{4}{c}{FOXA1}\\\cline{2-5}\cline{7-10}
$d$ & PICS & QuEST & MACS & CisGenome && PICS & QuEST & MACS & CisGenome\\\hline
$250$ & 188(73) & 405(63) & 0 & 0 && 6(83) & 269(67) & 0 & 0\\
$500$ & 376(71) & 950(63) & 0 & 0 && 26(70) & 361(68) & 0 & 0 \\
$1000$ & 478(70) & 1074(63) & 0 & 128(64) && 75(78) & 443(66) & 0 & 0\\
\hline
\end{tabular}
\end{center}
\end{table}

As described in section \ref{s:inf}, PICS can compute
approximate standard errors for its model parameter estimates. In
particular, we can derive an approximate confidence interval for a
given predicted binding event location as
$\hat{\mu}\pm c \cdot \mathrm{SE}(\hat{\mu})$, where $c$ is a constant to be
chosen as a function of the coverage desired. Assuming that
$\hat{\mu}$ is approximately normal, $c=1.96$ should give us an
approximate 95\% confidence interval for our binding site position.

\begin{figure}
 \centerline{\includegraphics[scale=0.6,angle=270]{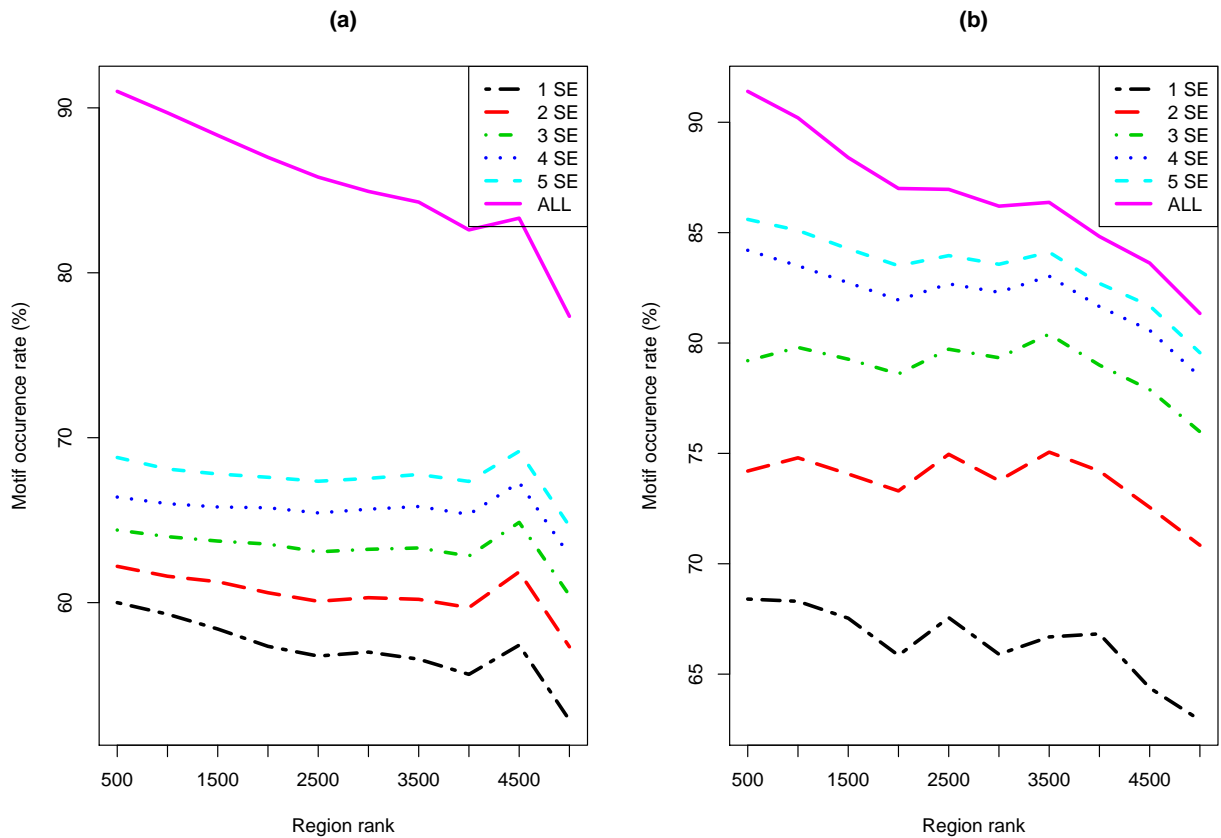}}
\caption{
The fraction of predicted binding events that had a GABP (a) or FOXA1 (b) motif site within  $c \cdot \mathrm{SE}(\mu$) of the predicted event location, $\mu$.} \label{f:quest.se}
\end{figure}

Using the set of motifs identified by GADEM, we evaluated
the actual coverage of our confidence intervals for different values
of $c$. Figure~\ref{f:quest.se} shows the occurrence frequency of
GABP motifs (left) and FOXA1 motifs (right) within $c \cdot \mathrm{SE}(\mu$)
of peaks centers. Using 3 standard errors, the coverage was approximately 65\% and 80\% for the GABP and FOXA1 data.
While these numbers suggest that the current version of
PICS provides a capable modeling framework, they also suggest that there are
significant opportunities to address noise and biases in more depth in order to improve spatial accuracy.

% Figure 6 - NEW CAPTION!

\begin{figure}
 \centerline{\includegraphics[scale=0.6,angle=270]{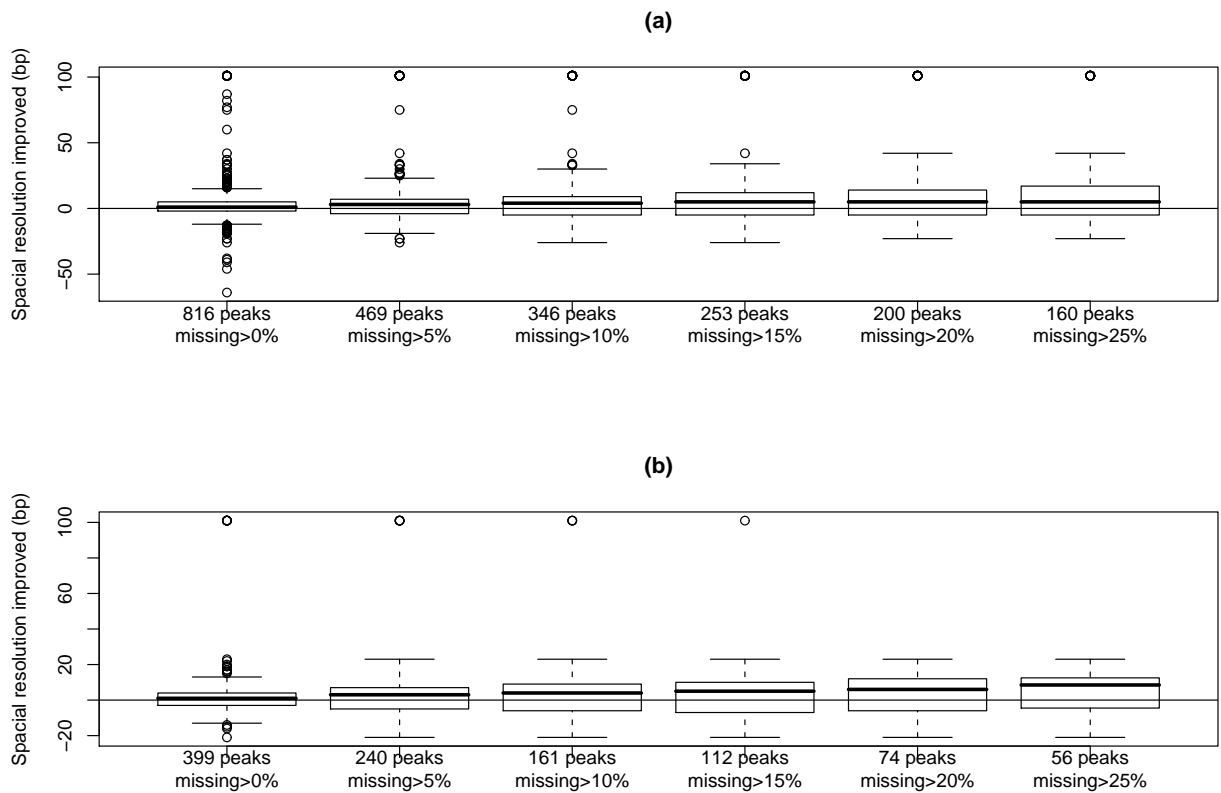}}
\caption{Using mappability improved spatial accuracy for (a) GABP and (b) FOXA1 data. The Y-axis shows how correcting for missing reads in predicting a binding site changed the distance between the site and the predicted binding motif closest to it. A positive (negative) value indicates that using the mappability correction decreased (increased) the distance between a site and its closest motif. For each data set, six relative levels of correction are shown, \emph{e.g.} for 200 GABP binding event regions, the final number of estimated reads in each region included at least 20\% of missing reads. 
} \label{f:box_map}
\end{figure}

Finally, we evaluated the effect of the mappability profiles on the parameter estimates. We
re-did the analysis while ignoring mappability, and compared the spatial accuracy, \emph{i.e.} the distance to the closest computationally verified binding site, with and without the mappability correction. Figure~\ref{f:box_map} shows boxplots of the difference
between corrected and uncorrected estimates for various percentage of missing reads.
The bloxplots are skewed to right, which shows that the correction improved the estimates for
binding event locations, and the degree of improvement increased with the fraction of missing reads.

% Figure 7

\section{Discussion}\label{s:discuss}

We have developed PICS, a probabilistic framework for detecting transcription
factor binding events from ChIP-seq experiments. The approach integrates a number 
of important factors in interpreting aligned read data, including correcting for 
reads that are missing due to genome repetitiveness and using prior 
information on input DNA fragment lengths.
Working with two published ChIP-seq data sets from human cell lines, we 
compared PICS to three alternative analysis methods. While additional methods are 
available for detecting bound regions from ChIP-seq 
\citep{b21,b504,b505,b507,b506}, the three methods we used 
have been shown to have good performance, and so offer 
reasonable performance baselines. The results of the comparison showed that, 
although the FDR-rank relationships returned differed by method and data set, the 
binding events predicted by PICS were the most consistent with computationally 
identified motif sites in both data sets.

We showed that PICS' mixture model
addresses multiple adjacent enrichment events, and can fit a
different DNA fragment length value for each binding event in a
mixture. While we allowed the mixture model to detect up to
15 components per candidate region, we can readily adjust this limit.
Datasets can be expected to contain regions in
which adjacent binding sites are too close to be resolved, but, given a
DNA fragment length distribution, we anticipate
that PICS should discriminate most adjacent sites that are
resolvable.

We note that, because it is based on mixture models and accounts for missing
reads, PICS is computationally intensive. The results shown were
obtained with an implementation of PICS that was written in the R
programming language \citep{b53}. Processing a $~$10M read data set required an
average computing time of three 3GHz CPU-hours per chromosome. While
we reduced the overall computation time by treating chromosomes in
parallel on a multiprocessor machine, and could also use a compute
cluster, we are also re-implementing PICS in C. We anticipate that
this new version will reduce the computing time by at least a factor
of ten and will scale well with larger datasets. PICS will be made freely available via Bioconductor \citep{b565}.

At the time of writing, all published short read ChIP-seq data are for single end (SE) reads, rather than for paired-end (PE) reads. PE data offer more direct information on DNA fragment lengths, should resolve a subset of read alignments that would be non-unique in SE data, and, in principle, could give direct information about long range chromosome interactions and genome rearrangements \citep{b1}. However, because a PE experiment  requires more input DNA and is more costly than an SE experiment, it is likely that PE and SE data will be appropriate for somewhat different applications. We anticipate that PICS will be useful in work to identify optimal applications for each approach, and that its probabilistic approach will remain useful for PE data, where having defined fragment lengths should simplify the modeling framework.

As a first step in implementing a probabilistic approach for ChIP-seq data, we have shown how to incorporate prior information about the DNA fragment lengths using a Bayesian approach. We can extend the PICS framework to incorporate more types of prior information. For example, we could place a prior distribution on $\mu$, the binding site position, and could include in this information about nucleosome occupancy and computationally derived motifs. Such extensions should allow us to further improve the detection of biologically relevant binding sites. With such extensions, we anticipate that probabilistic methods may help ChIP-seq contribute to biological research by offering principled ways for addressing backgrounds and diverse types of noise, and for integrating diverse types of biological information.

\section*{Acknowledgements}
We gratefully acknowledge Inanc Birol for discussions related to read mappability. We thank Martin Hirst, Anthony Fejes, Misha Bilenky and Nina Thiessen for suggestions that improved the manuscript. This research is supported by an NSERC Discovery Grant (RG and XZ).

\pagebreak

\appendix
\noindent\textbf{Computational details for the missing read case:}
We calculate expectations $\mathbb{E}_{dl}$ with respect to the double truncated
$t$-mixture density of unobserved reads as follows:
%\begin{eqnarray*}
%% \nonumber to remove numbering (before each equation)
%\mathbb{E}_{dl}( \tilde{\chi}_{dlk} |\tilde{n}_{dl}) &=& \tilde{n}_{dl}\mathbb{E}_{dl}[\tilde{z}_{dlk}]                                   = w_k P_{dl}^{-1}(\bTheta^-) H_{3,dlk}\\
%\mathbb{E}_{dl}( \tilde{s}_{dlk}    |\tilde{n}_{dl}) &=& \tilde{n}_{dl}\mathbb{E}_{dl}[\tilde{z}_{dlk}\tilde{u}_{dlk}]                    = w_k P_{dl}^{-1}(\bTheta^-) H_{0,dlk} \\
%\mathbb{E}_{dl}( \tilde{m}_{dlk}    |\tilde{n}_{dl}) &=& \tilde{n}_{dl}\mathbb{E}_{dl}[d_l\tilde{z}_{dlk}\tilde{u}_{dlk}]                 = w_k P_{dl}^{-1}(\bTheta^-) [2 \sigma^-_k H_{1,dlk}+ \mu^-_k H_{0,dlk}] \\
%\mathbb{E}_{dl}( \tilde{\eta}_{dlk} |\tilde{n}_{dl}) &=& \tilde{n}_{dl}\mathbb{E}_{dl}[(d_l-\hat{\mu}_k)^2\tilde{z}_{dlk}\tilde{u}_{dlk}]\\
%                                                    &=&  w_k P_{dl}^{-1}(\bTheta^-) [4 (\sigma_k^-)^2 H_{2,dlk}+(\mu^-_k-\hat{\mu}_k)^2H_{0,dlk}+ 4(\mu^-_k-\hat{\mu}_k)\sigma^-_kH_{1,dlk}]
%\end{eqnarray*}
\begin{eqnarray*}
\mathbb{E}_{dl}[\tilde{z}_{dlk}]                                   &=& w_k P_{dl}^{-1}(\bTheta^-) H_{3,dlk}\\
\mathbb{E}_{dl}[\tilde{z}_{dlk}\tilde{u}_{dlk}]                    &=& w_k P_{dl}^{-1}(\bTheta^-) H_{0,dlk} \\
\mathbb{E}_{dl}[d_l\tilde{z}_{dlk}\tilde{u}_{dlk}]                 &=& w_k P_{dl}^{-1}(\bTheta^-) [2 \sigma^-_k H_{1,dlk}+ \mu^-_k H_{0,dlk}] \\
\mathbb{E}_{dl}[(d_l-\hat{\mu}_k)^2\tilde{z}_{dlk}\tilde{u}_{dlk}] &=&  w_k P_{dl}^{-1}(\bTheta^-) [4 (\sigma_k^-)^2 H_{2,dlk}+(\mu^-_k-\hat{\mu}_k)^2H_{0,dlk}+ 4(\mu^-_k-\hat{\mu}_k)\sigma^-_kH_{1,dlk}]
\end{eqnarray*}

The quantities $H$'s can be calculated as:
\begin{eqnarray*}
  H_{j,dlk} &=&  h_j\left(\frac{b_l-\mu_{dk}}{2\sigma_{dk}})-h_j( \frac{a_l-\mu_{dk}}{2\sigma_{dk}}\right),\\
  H_{3,dlk} &=&  T_4(b_l|\mu_{dk},\sigma_{dk})-T_4(a_l|\mu_{dk},\sigma_{dk})
\end{eqnarray*}
for $j=0,1,2$, where $T_4$ refers to the c.d.f. of t distribution with
$4$ degrees of freedom, $B=\Gamma(3.5)/(\Gamma(3)\sqrt{\pi})$ is
a constant, and the functions $h_j$'s are defined as:
\begin{eqnarray*}
h_0(x)&=& B \left(\frac{1}{5} [h_4(x)]^5  -\frac{2}{3} [h_4(x)]^3  + h_4(x) \right)\\
h_1(x)&=& B \left(\frac{1}{3} [h_4(x)]^3  -\frac{1}{5} [h_4(x)]^5  \right)\\
h_2(x)&=& \frac{-B}{5}\left(1+x^2\right)^{-2.5}\\
h_4(x)&=& \sin (\arctan (x))
\end{eqnarray*}

\noindent\textbf{Parameter recalculation when merging binding events:}
The parameters of merged binding events are calculated by solving these moment matching equations:
\begin{eqnarray*}
  \overline{\mu} &=& \frac{\sum_k \mu_k w_k}{ \sum_k w_k} \\
  \overline{\delta} &=& \frac{\sum_k \delta_k w_k}{ \sum_k w_k} \\
  \frac{\nu}{\nu-2} \overline{\sigma_f^2} - (\overline{\mu}- \frac{\overline{\delta}}{2})^2 &=& \sum_k\left[w_k (\frac{\nu}{\nu-2} \sigma_fk^2 - (\mu_k- \frac{\delta_k}{2})^2)\right]\\
  \frac{\nu}{\nu-2} \overline{\sigma_r^2} - (\overline{\mu}+ \frac{\overline{\delta}}{2})^2 &=& \sum_k\left[w_k (\frac{\nu}{\nu-2} \sigma_rk^2 - (\mu_k+ \frac{\delta_k}{2})^2)\right].
\end{eqnarray*}

\label{lastpage}

\end{document}